\documentclass[journal]{IEEEtran}

\usepackage{amssymb}
\usepackage{amsmath}
\usepackage{multirow}
\usepackage{tabularx}
\usepackage[pdftex]{graphicx}
\graphicspath{{figures/}}
\usepackage{subfig}
\usepackage{url}
\usepackage{cite}
\begin{document}
\bstctlcite{BSTcontrol}

\title{Modelling the Nonlinear Response\\ of Silicon Photomultipliers}

\author{Jaime~Rosado%
\thanks{J. Rosado is with the IPARCOS Institute and the EMFTEL Department, Universidad Complutense de Madrid, E-28040 Madrid, Spain (e-mail: jrosadov@ucm.es).}%
\thanks{\textcopyright 2019 IEEE. Personal use of this material is permitted. Permission from IEEE must be obtained for all other uses, in any current or future media, including reprinting/republishing this material for advertising or promotional purposes, creating new collective works, for resale or redistribution to servers or lists, or reuse of any copyrighted component of this work in other works.}}

\maketitle

\begin{abstract}
A statistical model of the nonlinear response of silicon photomultipliers is presented. It includes losses of both the photodetection efficiency and the gain during pixel recovery periods as well as the effect of correlated and uncorrelated noise. The model provides either the mean output charge of a SiPM for incident light pulses of arbitrary shape or the output current for continuous light. The dependence of the SiPM response on both the overvoltage and the pulse shape is also properly described.
The model has been validated for two different silicon photomultipliers using scintillation light pulses from a LYSO crystal as well as continuous light from a LED. Good agreement is found with experimental data at moderate nonlinearity.

\end{abstract}


\section{Introduction}
\label{sec:intro}

\IEEEPARstart{T}{he} silicon photomultiplier (SiPM) is a high-sensitivity semiconductor photodetector that is increasingly becoming the best choice in many applications owing to its compactness, high gain, insensitivity to magnetic fields, rapid response and exceptional photon-counting resolution. This device consists in an array of Geiger-mode avalanche photodiodes in the $\mu$m scale, hereafter called pixels, which are sensitive to single photons. The pixels are connected in parallel in such a way the SiPM output signal is the sum of the signals from all the pixels. In this work, the SiPM response for an incident light pulse is characterized by the mean output charge, which is given by $\langle Q\rangle=\varepsilon\cdot n\cdot q$ for an ideal SiPM, where $\varepsilon$ is the photodetection efficiency, $n$ is the mean number of photons per pulse impinging on the SiPM surface and $q$ is the mean charge released by a single breakdown avalanche, which characterizes the gain of the device. 

The mean avalanche charge $q$ is proportional to the overvoltage $U$, which is defined as the reverse-bias voltage $V_{\rm bias}$ minus the breakdown voltage $V_{\rm br}$ of the SiPM. Both $q$ and $V_{\rm br}$ can be obtained from the output pulse charge spectrum at photon counting levels (see, e.g., \cite{Rosado2015}). For typical blue sensitive SiPMs based on a p-on-n structure, $\varepsilon$ is described by an exponential function of $U$ \cite{Otte2017}. In this work it is assumed
\begin{equation}
\label{eq:varepsilon}
\varepsilon=\varepsilon_{\rm max}\cdot\left[1-\exp\left(-\frac{U-U_0}{U_{\rm ch}}\right)\right]
\end{equation}
for $U\geq U_0$, otherwise $\varepsilon=0$. Here $\varepsilon_{\rm max}$ is the product of the quantum efficiency and the fill factor (i.e., the ratio of the photosensitive area of a pixel to the entire pixel area), the parameter $U_{\rm ch}$ is a characteristic overvoltage of the SiPM depending on the input wavelength spectrum \cite{Otte2017}, and $U_0$ is a small overvoltage shift that is introduced to better fit experimental data of $\varepsilon$ (see \cite{Nagy2017,Chmill2017} for an interesting studies on the breakdown behaviour of SiPMs). It should be clarified that this empirical expression is not applicable to n-on-p SiPMs \cite{Gallina2019}. 

The SiPM response is nonlinear for high photon densities because the number of pixels is finite (typically of the order of 1000, depending on the SiPM) and they take a few tens of ns to recharge after each breakdown avalanche. Besides, SiPMs have uncorrelated noise due to the thermal production of electron-hole pairs in silicon, as well as correlated noise caused by two processes called afterpulsing and crosstalk. Afterpulses are stochastic parasitic avalanches produced in the same pixel where a previous avalanche has been triggered. On the other hand, crosstalk avalanches are produced in nearby pixels either simultaneously with the primary avalanche (prompt crosstalk) or with some delay (delayed crosstalk) \cite{Rosado2015}.

The combination of all these features makes the statistics of avalanche triggering and charge production very complex. As a consequence, an exact statistical description of the response of SiPMs is only possible via Monte Carlo simulations. H.T. van Dam \emph{et al.} \cite{VanDam2010} developed a comprehensive statistical model of the SiPM response for exponentially decaying light pulses, accounting for both the correlated noise and the pixel recovery under some approximations. However it requires non-trivial numerical calculations. A simpler statistical model was presented in \cite{Jeans2016}, but it ignores the correlated noise and accounts of the recovery of pixels in an incomplete way. In addition, analytical expressions of the SiPM response in two limit situations are available. First, for light pulses much shorter than the pixel recovery time and in the absence of correlated and uncorrelated noise, the mean output charge $\langle Q\rangle$ is given by the well-known expression (see, e.g., \cite{Renker2006})
\begin{equation}
\label{eq:InstPulse}
\langle Q\rangle=N\cdot q\cdot\left[1-\exp\left(-\frac{\varepsilon\cdot n}{N}\right)\right]\,,
\end{equation}
where $N$ is the number of pixels. Second, for continuous light and ignoring again the correlated noise, the output current intensity $I$ produced by the SiPM can be approximated by
\begin{equation}
\label{eq:NonParalyzable}
I=\frac{\varepsilon\cdot r\cdot q}{1+\varepsilon\cdot r\cdot t_{\rm dead}/N}\,,
\end{equation}
where $r$ is the rate of impinging photons and $t_{\rm dead}$ is a certain non-paralyzable dead time accounting for the pixel recovery \cite{Vinogradov2015a}. Modified versions of the above expressions are being used to describe the response of SiPMs at particular conditions, e.g., \cite{Kotera2016,Vinogradov2015b,Grodzicka2015,Niu2012,Pulko2012}. However, a general analytical model is not available yet.

In this paper, a simple model that includes all the above features under some simplifications is described. A preliminary version of this model was first reported in \cite{Rosado2018}, but significant upgrades have been made since then. The model has been validated against experimental data for scintillation light pulses from a LYSO crystal as well as for continuous light from a LED.

\section{The model}
\label{sec:model}

\subsection{Approach of the problem}
\label{ssec:approach}

In steady state, all the pixels of a SiPM are biased with the supplied overvoltage $U$. Therefore, the probability that an impinging photon fires a pixel is equal to the nominal photodetection efficiency $\varepsilon$ given by (\ref{eq:varepsilon}). If the photon is detected, the mean output charge of the SiPM is equal to $q$ in the absence of correlated noise.

When a breakdown avalanche is triggered in a pixel, the instantaneous overvoltage $u$ of that pixel drops to zero and then grows exponentially as
\begin{equation}
\label{eq:u_vs_t}
u(t)=U\cdot\left[1-\exp\left(-\frac{t}{t_{\rm rec}}\right)\right]\,,
\end{equation}
where $t$ is the delay time from the avalanche triggering and $t_{\rm rec}$ is the recovery time of the SiPM, which ranges from tens of ns to several $\mu$s depending on the device. During the recovery period, both the probability that a photon triggers a new avalanche in the pixel and the mean charge released by this avalanche (if triggered) depend on $u$. The recovery functions of the photodetection efficiency and the gain can be described by
\begin{align}
\varepsilon_{\rm rec}(t)&=\varepsilon_{\rm max}\cdot\left[1-\exp\left(-\frac{u(t)-U_0}{U_{\rm ch}}\right)\right] \label{eq:a}\\
q_{\rm rec}(t)&=q\cdot\frac{u(t)}{U} \label{eq:b}
\end{align}
for $t>t_0$, where the time shift $t_0$ is defined by $u(t_0)=U_0$. For $t<t_0$, $\varepsilon_{\rm rec}(t)=0$ and $q_{\rm rec}(t)=0$. The time shift $t_0$ can generally be assumed to be much lower than $t_{\rm rec}$. However, $t_0$ may become significant when $U$ approaches $U_0$, making $\varepsilon_{\rm rec}(t)$ to grow more slowly than $q_{\rm rec}(t)$.

If several photons hit the same pixel, the time evolution of avalanche triggering and pixel recovery may be very complicated, especially if the light pulse duration is comparable to $t_{\rm rec}$. Besides, each avalanche can induce correlated noise that also contributes to the total output charge. The probabilities of crosstalk and afterpulsing are proportional to the number of charge carriers in the primary avalanche and also depend on the instantaneous overvoltages in the pixels where these secondary processes may take place. Taking into account that secondary avalanches can in turn induce correlated noise, this results in a complex chain process that depends on the temporal and spatial characteristics of crosstalk and afterpulsing (see \cite{Gallego2013,Rosado2015} for details).

Despite of the complexity of the problem, the whole process of charge production for each impinging photon can be modelled in three steps:

i) The photon has a probability $\varepsilon$ to produce an \emph{avalanche seed} in the pixel that it hits. This event is independent of whether or not more seeds are produced in the same or other pixels.
    
ii) If the seed is produced, it has a probability $\varepsilon_{\rm rec}(t_s)/\varepsilon$ to develop into an avalanche with mean charge $q_{\rm rec}(t_s)$, where $t_s$ is the time difference between the occurrence of this seed $s$ and the last avalanche triggered in the same pixel. Notice that $\varepsilon_{\rm rec}(t_s)/\varepsilon=1$ and $q_{\rm rec}(t_s)=q$ for $t_s\gg t_{\rm rec}$, which corresponds to the case that the seed is produced in a pixel with $u=U$.
        
iii) If the avalanche is triggered, it may induce crosstalk and afterpulsing. For simplicity, these processes are not considered to produce further seeds, but the total charge from the primary avalanche and all the ensuing secondary ones triggered in the same or other pixels is associated to the same seed. The net charge $q_s$ from this seed will have a certain probability density function $f_s(t_s,q_s)$ with mean $q_{\rm rec}(t_s)\cdot[1+c_s(t_s)]$, where $c_s(t_s)$ stands for the mean value of the relative contribution from secondary avalanches to $q_s$. Notice that, besides the explicit dependence on $t_s$, $f_s(t_s,q_s)$ and $c_s(t_s)$ are generally different for each seed because they depend on the occurrence of both primary and secondary avalanches from all other seeds produced during the light pulse. In case that only one seed is produced, the probability density function and mean value of the net seed charge are simpler and will be referred to respectively as $f(q_s)$ and $q\cdot(1+c)$.

The number of seeds $m$ produced by a light pulse has a Poisson distribution
\begin{equation}
\label{eq:ProbPoiss}
P(m)=\frac{(\varepsilon\cdot n)^m}{m!}\cdot e^{-\varepsilon\cdot n}\,,
\end{equation}
with $\varepsilon\cdot n$ being the mean number of seeds per light pulse. On the other hand, the total output charge $Q=\sum_s q_s$ depends on the correlations between the avalanche triggering probabilities and the net charges of seeds. These correlations can be understood as the cause of nonlinearity of the SiPM response. Under this scheme, the probability density function of $Q$ can be expressed as
\begin{equation}
\label{eq:FQ}
F(Q)=\sum_{m=0}^\infty P(m)\cdot F_m(Q)\,,
\end{equation}
where $F_m(Q)$ is the probability density function of $Q$ in the case that exactly $m$ seeds are produced. Then, the expectation value of $Q$ is
\begin{equation}
\label{eq:MeanQ}
\langle Q\rangle=\sum_{m=0}^\infty P(m)\cdot \langle Q\rangle_m\,,
\end{equation}
where
\begin{equation}
\label{eq:MeanQm}
\langle Q\rangle_m=\int_0^\infty F_m(Q)\cdot Q\cdot{\rm d}Q\,.
\end{equation}

In the case that only one seed is produced (i.e., $m=1$), $Q$ is equal to the net charge $q_s$ of the seed, hence $F_1(Q)=f(q_s)$ and $\langle Q\rangle_1=q\cdot(1+c)$. However, even for this simple case, a rigorous theoretical determination of $c$ would require to describe the development of chains of secondary avalanches, including afterpulses that induce crosstalk, and vice versa. In addition, measurements of all the relevant parameters for the calculation of $c$ are arduous and subjected to large uncertainties (see, e.g., \cite{Rosado2015}). For this reason, $c$ is taken as a free model parameter that is determined from data in the linear response regime of the SiPM, as will be shown later.

Notice that all the complex dynamics of pixel recovery and the interactions between seeds are enclosed in the functions $F_m(Q)$. In the following, $F_m(Q)$ and $\langle Q\rangle_m$ are derived for an arbitrary number of seeds $m$ under some approximations.

\subsection{Exact solution for a simple case}
\label{ssec:two_seeds}

Before going to the general situation, let us consider first $m=2$ and no correlated noise. The two seeds are generally produced in two different pixels with $u=U$. In this case, both seeds develop into two independent avalanches with charge distributions $f(q_1)$ and $f(q_2)$, because they do not interact with each other in the absence of correlated noise. In addition, the mean net charge is $q$ for each seed because $c=0$. Therefore, the probability density function of the total output charge is given by
\begin{equation}
\label{eq:F2a}
F_2(Q)=\int_0^Q f(Q-q_s)\cdot f(q_s)\cdot{\rm d}q_s\,, 
\end{equation}
where the integration over $q_s$ accounts for all the possible combinations of $q_1$ and $q_2$ adding up to $Q$. The mean output charge is simply $\langle Q\rangle_2= 2\cdot q$.

However, in the unlikely event that both seeds are produced in the same pixel at times $t'$ and $t''$, the first one still develops into an avalanche with charge density function $f(q_1)$ and mean charge $q$, whereas the second one has a probability $\varepsilon_{\rm rec}(t_2)/\varepsilon$ to trigger an avalanche, provided that $t_2=t''-t'>t_0$. If triggered, the charge distribution of this second avalanche has a probability density function $f_2(t_2,q_2)$ with mean $q_{\rm rec}(t_2)$ in the absence of correlated noise.

Now, let us consider that the two seeds are produced in the same pixel randomly with a time distribution $p(t)$, which corresponds to the distribution of the arrival times of photons. The probability that both seeds develop into an avalanche is
\begin{equation}
\label{eq:alpha}
\alpha=\frac{2}{\varepsilon}\cdot\int_0^\infty \int_{t_0}^\infty p(t)\cdot p(t+t_s)\cdot\varepsilon_{\rm rec}(t_s)\cdot{\rm d}t_s\cdot{\rm d}t\,,
\end{equation}
where the variables of integration are $t=\min(t',t'')$ and $t_s=\lvert t'-t''\rvert$, and the factor 2 accounts for both possibilities $t'<t''$ and $t'>t''$. Here it is assumed $p(t)>0$ for $t>t_0$, otherwise $\alpha=0$.

Similarly, it can be defined a time-averaged charge density function for the second seed $s$ (whichever it is) as
\begin{multline}
\label{eq:phi}
\phi(q_s)=\frac{2}{\alpha\cdot\varepsilon}\cdot\int_0^\infty \int_{t_0}^\infty p(t)\cdot p(t+t_s)\\
\cdot\varepsilon_{\rm rec}(t_s)\cdot f_s(t_s,q_s)\cdot{\rm d}t_s\cdot{\rm d}t\,,
\end{multline}
which is properly normalized so that $\int_0^\infty\phi(q_s)\cdot{\rm d}q_s=1$. The mean of this distribution can be expressed as $\gamma\cdot q/\alpha$, where the dimensionless parameter $\gamma$ is defined by
\begin{equation}
\label{eq:gamma}
\gamma=\frac{2}{\varepsilon\cdot q}\cdot\int_0^\infty \int_{t_0}^\infty p(t)\cdot p(t+t_s)\cdot\varepsilon_{\rm rec}(t_s)\cdot q_{\rm rec}(t_s)\cdot{\rm d}t_s\cdot{\rm d}t\,.
\end{equation}

From this result, the probability density function of the sum charge for two seeds uniformly distributed over the SiPM area and with time distribution $p(t)$ in the absence of correlated noise is
\begin{multline}
\label{eq:F2b}
F_2(Q)=\int_0^Q \left(1-\frac{1}{N}\right)\cdot f(Q-q_s)\cdot f(q_s)\cdot {\rm d}q_s\\
      +\frac{1-\alpha}{N}\cdot f(Q)+\int_0^Q\frac{\alpha}{N}\cdot f(Q-q_s)\cdot\phi(q_s)\cdot{\rm d}q_s\,.
\end{multline}
The three terms of the right-hand side of this equation arise from the following contributions: i) the two seeds develop into two avalanches in different pixels, ii) both seeds are produced in the same pixel and only the first one develops into an avalanche, iii) both seeds are produced in the same pixel and develop into two avalanches.

Substituting (\ref{eq:F2b}) into (\ref{eq:MeanQm}) results in
\begin{equation}
\label{eq:Q2}
\langle Q\rangle_2=2\cdot q\cdot\left(1-\frac{1}{N}\right)+\frac{(1+\gamma)\cdot q}{N}\,,
\end{equation}

\subsection{Approximations for the general case}
\label{ssec:general}

It becomes unwieldy to use the above procedure for arbitrary $m$ and including both correlated and uncorrelated noise. Nevertheless, (\ref{eq:F2b}) and (\ref{eq:Q2}) can be generalized under some simplifications:

i) For a total output charge $Q$, the mean number of fired pixels is $Q/q$ to first-order approximation, which is equivalent to assume that each avalanche is triggered in a different pixel and has charge $q$. In reality, several avalanches (both primary and secondary ones) may be produced in the same pixel, but this is partly compensated by the fact that the greater the number of avalanches in a pixel, the smaller the mean charge per avalanche as a consequence of the reduction of the pixel gain during recovery periods. Under this approximation, the probability that a given seed $s$ is produced in a \emph{busy pixel} is $(Q-q_s)/(N\cdot q)$, where $q_s$ is the net charge contribution (if any) from this seed and $Q-q_s$ is that from all other seeds. Otherwise, the seed is considered to be produced in a \emph{free pixel} with $u=U$.

ii) If the seed is produced in a free pixel, it is assumed that the avalanche triggering probability of the seed is unity and that the ensuing secondary avalanches are unaffected by the presence of other seeds. Therefore, the probability density distribution and the expectation value of $q_s$ are $f(q_s)$ and $q\cdot(1+c)$, respectively, where $c>0$ now. Notice that, although some seed interactions are ignored under this approximation, the correlated noise contributes to the nonlinearity via a reduction in the number of free pixels.

iii) If the seed is produced in a busy pixel, the probability that an additional avalanche is triggered is assumed to be equal to $\alpha$, as given by (\ref{eq:alpha}), that is, only interactions between two seeds are considered. The distribution of the net charge of the last seed (provided that it develops into an avalanche) is still referred to as $\phi(q_s)$, which is defined by (\ref{eq:phi}) but with the caveat that it should include the charge contribution from correlated noise. Assuming again that secondary avalanches are unaffected by the presence of other seeds, the mean net charge of the seed is $\gamma\cdot q\cdot(1+c)/\alpha$ in this case.

iv) The function $p(t)$ needed to calculate $\alpha$ and $\gamma$ from (\ref{eq:alpha}) and (\ref{eq:gamma}) is no longer the distribution of the arrival times of photons, because the actual time distribution of avalanches depends on both the temporal characteristics of the correlated noise and the interactions between seeds. Instead, $p(t)$ is approximated by the normalized output pulse signal, since it does include these effects. Note, however, that it may be necessary to perform deconvolution if the output pulse is of similar length to the single-photon response function. In appendices \ref{sec:rectangular} and \ref{sec:exponential}, useful expressions for $\gamma$ are derived for typical pulse shapes, namely rectangular and double exponential pulses.

v) The effect of uncorrelated noise in the SiPM response is approximately included following the procedure described in appendix \ref{sec:uncorrelated}. However, the corrections to be made do not affect the derivation and the interpretation of the main features of the model.

From the above considerations it follows this recurrence equation for $m>1$:
\begin{align}
\label{eq:Fm}
F_m(Q)=&\int_0^Q\left(1-\frac{Q-q_s}{N\cdot q}\right)\cdot F_{m-1}(Q-q_s)\cdot f(q_s)\cdot{\rm d}q_s \nonumber\\ 
&+\frac{(1-\alpha)\cdot Q}{N\cdot q}\cdot F_{m-1}(Q)\\
&+\int_0^Q \frac{\alpha\cdot(Q-q_s)}{N\cdot q}\cdot F_{m-1}(Q-q_s)\cdot\phi(q_s)\cdot{\rm d}q_s\,, \nonumber
\end{align}
where the three terms of the right-hand side have a similar meaning to those of (\ref{eq:F2b}). This equation provides the probability that the last seed of the sequence is produced in either a free or busy pixel and its contribution to the total output charge. Substituting (\ref{eq:Fm}) into (\ref{eq:MeanQm}) leads to
\begin{equation}
\label{eq:Qm}
\langle Q\rangle_m=\frac{N\cdot q}{1-\gamma}\cdot\left[1-\left(1-\frac{(1-\gamma)\cdot(1+c)}{N}\right)^m\right]\,,
\end{equation}

\subsection{Mean response to light pulses}
\label{ssec:MeanResponse}

\begin{figure}[!t]
\centering
\includegraphics[width=\linewidth]{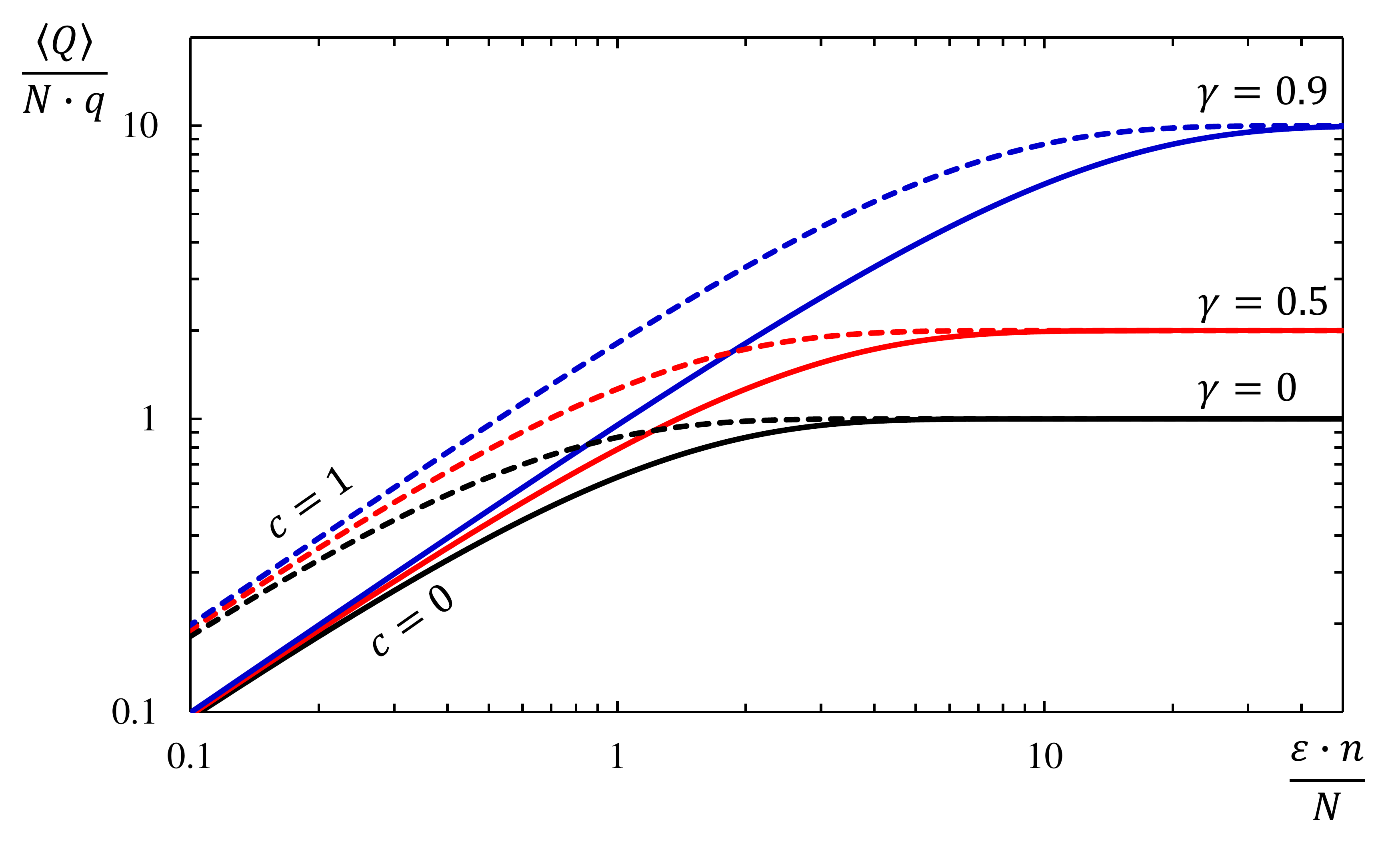}
\caption{Results from (\ref{eq:MeanQ_Poiss}) for different combinations of $\gamma$ and $c$. The case of $\gamma=0$ and $c=0$ (black solid line) corresponds to (\ref{eq:InstPulse}) for instantaneous light pulses and no correlated noise.}
\label{fig:MeanQ}
\end{figure}

The above result in combination with (\ref{eq:MeanQ}) leads to the following simple expression for the mean output charge of a SiPM for light pulses with a mean number $n$ of photons:
\begin{equation}
\label{eq:MeanQ_Poiss}
\langle Q\rangle=\frac{N\cdot q}{1-\gamma}\cdot\left[1-\exp\left(-(1-\gamma)\cdot(1+c)\cdot\frac{\varepsilon\cdot n}{N}\right)\right]\,.
\end{equation}
In Fig. \ref{fig:MeanQ}, the properties of this expression are illustrated in a plot of the normalized mean charge $\langle Q\rangle/(N\cdot q)$ versus $(\varepsilon\cdot n)/N$ for different values of $c$ and $\gamma$.

For $(\varepsilon\cdot n)/N\ll1$, the normalized mean charge increases linearly with $(\varepsilon\cdot n)/N$ with a slope $1+c$. In most practical cases, the input light intensity is however characterized by a quantity $X$ proportional to $n$ (e.g., the energy deposited in a calorimeter). Therefore, the factor $1+c$ can be included in a coefficient $k_X$ such that $\langle Q\rangle\approx\varepsilon\cdot q\cdot k_X\cdot X$ for $k_X\cdot X\ll 1$. If $\varepsilon$ and $q$ are known, $k_X$ can be determined from data in the linear response region of the SiPM. It should be clarified that, although the integration time window is large enough to contain the full output pulse, the measured value of $\langle Q\rangle$ may not include small contributions from afterpulses with long delay (see, e.g., \cite{Rosado2015,Vacheret2011}).

When $(\varepsilon\cdot n)/N\rightarrow\infty$, the normalized mean charge approaches $1/(1-\gamma)$, where $\gamma$ is calculated from (\ref{eq:gamma}) and quantifies the average charge losses due to interactions between seeds produced in the same pixel. For a pulse much longer than the recovery time, $\gamma$ approaches one. This makes that (\ref{eq:MeanQ_Poiss}) is approximately linear even for $(\varepsilon\cdot n)/N>1$, since each pixel may be fired and recovered several times during the pulse. In the opposite case of a very short pulse (i.e., the deconvolution of the output pulse has a length much shorter than $t_{\rm rec}$), one gets $\gamma\approx 0$ and the saturation level is unity. If $c=0$ can be assumed too, then (\ref{eq:MeanQ_Poiss}) reduces to (\ref{eq:InstPulse}), which corresponds to the limit situation for an instantaneous input light pulse and no correlated noise.

The model can also be adapted to describe the SiPM response to continuous light by assuming a rectangular pulse of length $T\gg t_{\rm rec}$ (see appendix \ref{sec:rectangular}) and making the substitutions $r=n/T$ and $I=\langle Q\rangle/T$, where $r$ is the incident photon rate and $I$ is the output current. This results in
\begin{equation}
\label{eq:I_cont}
I=\frac{N\cdot q}{2\cdot t_{\rm dead}}\cdot\left[1-\exp\left(-2\cdot t_{\rm dead}\cdot(1+c)\cdot\frac{\varepsilon\cdot r}{N}\right)\right]\,.
\end{equation}
where $t_{\rm dead}$ is defined by (\ref{eq:t_dead}) and it can be interpreted as an effective dead time. When $U$ is large enough so that $t_0\ll t_{\rm rec}$ and $\varepsilon\approx\varepsilon_{\rm max}$, then $t_{\rm dead}$ approaches $t_{\rm rec}$. Notice that, in the absence of correlated noise (i.e., $c=0$), the expansion of (\ref{eq:I_cont}) to first order in $(\varepsilon\cdot r\cdot t_{\rm dead})/N$ is equivalent to (\ref{eq:NonParalyzable}) for a non-paralyzable detector with dead time $t_{\rm dead}$.

\section{Experimental verification}
\label{sec:experiment}

The model was validated against experimental data for two SiPMs of the S13360 series from Hamamatsu, namely the type numbers S13360-1325CS and the S13360-1350CS \cite{Hamamatsu}. These detectors were chosen to have the same area but different pixel size, in such a way that nonlinearity is stronger in the 1350 SiPM (pixel size of 50 $\mu$m and $N=667$) than in the 1325 SiPM (pixel size of 25 $\mu$m and $N=2668$). The characteristics of these SiPMs are summarized in table \ref{tab:SiPMs}. The breakdown voltage $V_{\rm br}$, the dark count rate $r_{\rm dc}$ and the mean avalanche charge $q$ are those given in the datasheet of the devices for the recommended operating voltage $U_{\rm op}$ and at a reference temperature of $25^\circ$C. The parameters $U_{\rm ch}$, $U_0$ and $\varepsilon_{\rm max}$ were obtained by fitting (\ref{eq:varepsilon}) to the data of the photodetection efficiency in the range 2 V$< U<$ 9 V for 450 nm photons and at $25^\circ$C also available in the datasheet. Fit residuals were lower than 3\%. The values of the recovery time $t_{\rm rec}$ and the overall probability of correlated noise at $U_{\rm op}$ were taken from \cite{Rosado2015}.

\begin{table}[!t]
\caption{Characteristics of the Tested SiPMs.}%
\label{tab:SiPMs}
\centering%
\resizebox{\linewidth}{!}{
\begin{tabular}{|l|c|c|c|c|}
\hline
Parameters          & Symbol                  & 1325    & 1350    & Unit           \\
\hline
SiPM area           & -                       & \multicolumn{2}{c|}{$1.3\times1.3$} & mm$^2$         \\
\hline
Pixel pitch         & -                       & 25               & 50               & $\mu$m         \\
\hline
Number of pixels    & $N$                     & 2668             & 667              & -              \\
\hline
Breakdown voltage & $V_{\rm br}$            & \multicolumn{2}{c|}{51.80}          & V              \\
\hline
Recom. op. overvoltage & $U_{\rm op}$        & 5.00             & 3.00             & V              \\
\hline
Dark count rate     & $r_{\rm dc}$            & $<210$           & $<270$           & kcps           \\
\hline
Avalanche charge & $q$                     & 0.7              & 1.7              & $\times10^6$ e \\
\hline
Charact. overvoltage & $U_{\rm ch}$            & 2.69             & 2.68             & V              \\
\hline
Overvoltage shift   & $U_0$                   & 0.66             & 0.00             & V              \\
\hline
Maximum PDE         & $\varepsilon_{\rm max}$ & 0.327            & 0.597            & -              \\
\hline
Recovery time       & $t_{\rm rec}$           & 17               & 29               & ns             \\
\hline
Cor. noise probability  & -                   & 0.03             & 0.01            & -              \\
\hline
\end{tabular}}
\end{table}

A detailed experimental characterization of both SiPMs was previously reported in \cite{Rosado2015}. After correcting for the amplification factor used in those measurements, the values of $q$ obtained from the pulse charge spectra at photon counting levels are consistent within 10\% with the ones given in the datasheet. The correlated noise is dominated by afterpulsing and delayed crosstalk in both SiPMs, which have optical barriers between pixels that drastically reduce prompt crosstalk. The sum of the probabilities of the three components of correlated noise at $U_{\rm op}$ is shown in the table as a reference value, but the charge contribution of the correlated is effectively characterized by the parameter $c$ in this model.

The SiPMs were biased using a Hamamatsu C12332 driver circuit that incorporates a compensation system for the temperature dependence of the SiPM gain. The output signal was registered with a digital oscilloscope Tektronix TDS5032B with math functions, including pulse integration and histogramming. Two types of light sources were used. First, the SiPM was coupled to a scintillation crystal irradiated by $\gamma$ rays. Second, the SiPM was illuminated with continuous light from a LED.

\subsection{Scintillation light pulses}
\label{ssec:scintillation}

\begin{figure}[!t]
\centering
\includegraphics[width=\linewidth]{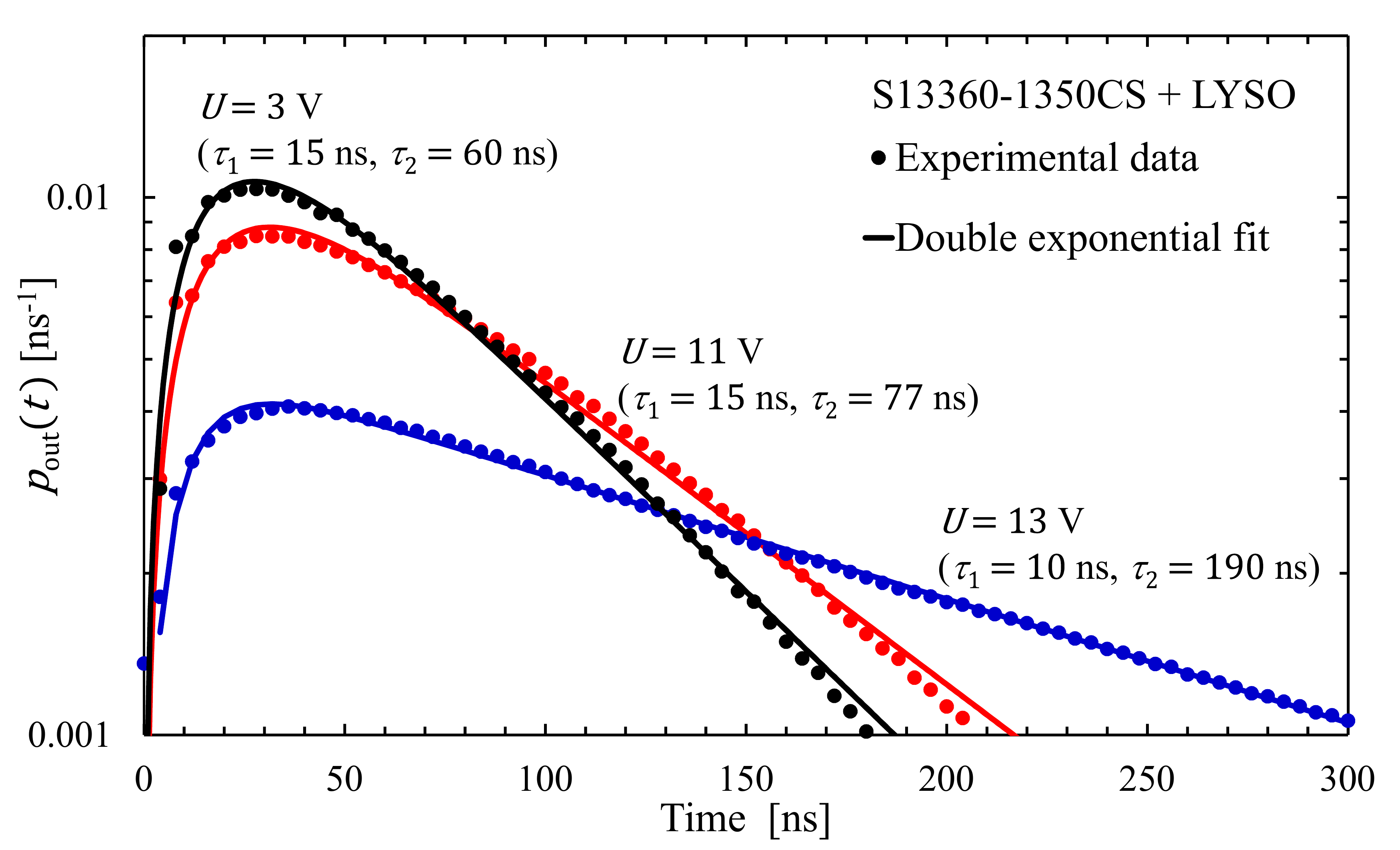}
\caption{Normalized output pulse shapes of the S13360-1350CS SiPM for LYSO scintillation pulses at the recommended overvoltage value $U=3$ V as well as at two very high overvoltage values, where the correlated noise becomes very significant. Lines represent the best fits to data by a double exponential function (\ref{eq:exponential}).}
\label{fig:pout}
\end{figure}

A LYSO(Ce) scintillation crystal of $3\times3\times20$ mm$^3$ with BaSO$_4$ reflector was used. The decay time and light yield of the LYSO(Ce) material are 42 ns and 29 photons/keV, respectively, according to the manufacturer's data report \cite{EpicCrystal}. Silicone grease was used for the optical coupling to the SiPM. Different radioactive sources ($^{22}$Na, $^{60}$Co, $^{137}$Cs and $^{226}$Ra) were used, providing a range of $\gamma$-ray energies from 300 to 2100~keV. An integration time window of 400 ns was set on the oscilloscope, which was large enough to contain the full output pulse (see Fig. \ref{fig:pout}), but still small so that contributions from uncorrelated noise can be neglected ($r_{\rm dc}\cdot T<0.1$).

Output charge spectra for the different radioactive sources were obtained at several overvoltage values for both SiPMs. The optimal values of the output charge resolution for the $^{137}$Cs photopeak at 662 keV were 13\%(FWHM) for the 1325 SiPM at $U=13$ V and 7\%(FWHM) for the 1350 SiPM at $U=5$ V, although the energy resolution was somewhat worse owing to the nonlinear response of the SiPMs, as discussed in \cite{Rosado2018}. In particular, the resolution degraded significantly at the highest energies for the 1350 SiPM. Nevertheless, seven photopeaks were clearly resolved in the complex spectrum of $^{226}$Ra for all $U$ values, except for the photopeaks at 295 keV and 351 keV, which were not well distinguished from the Compton continuum in some cases.

\begin{figure}[!t]
\centering
\includegraphics[width=\linewidth]{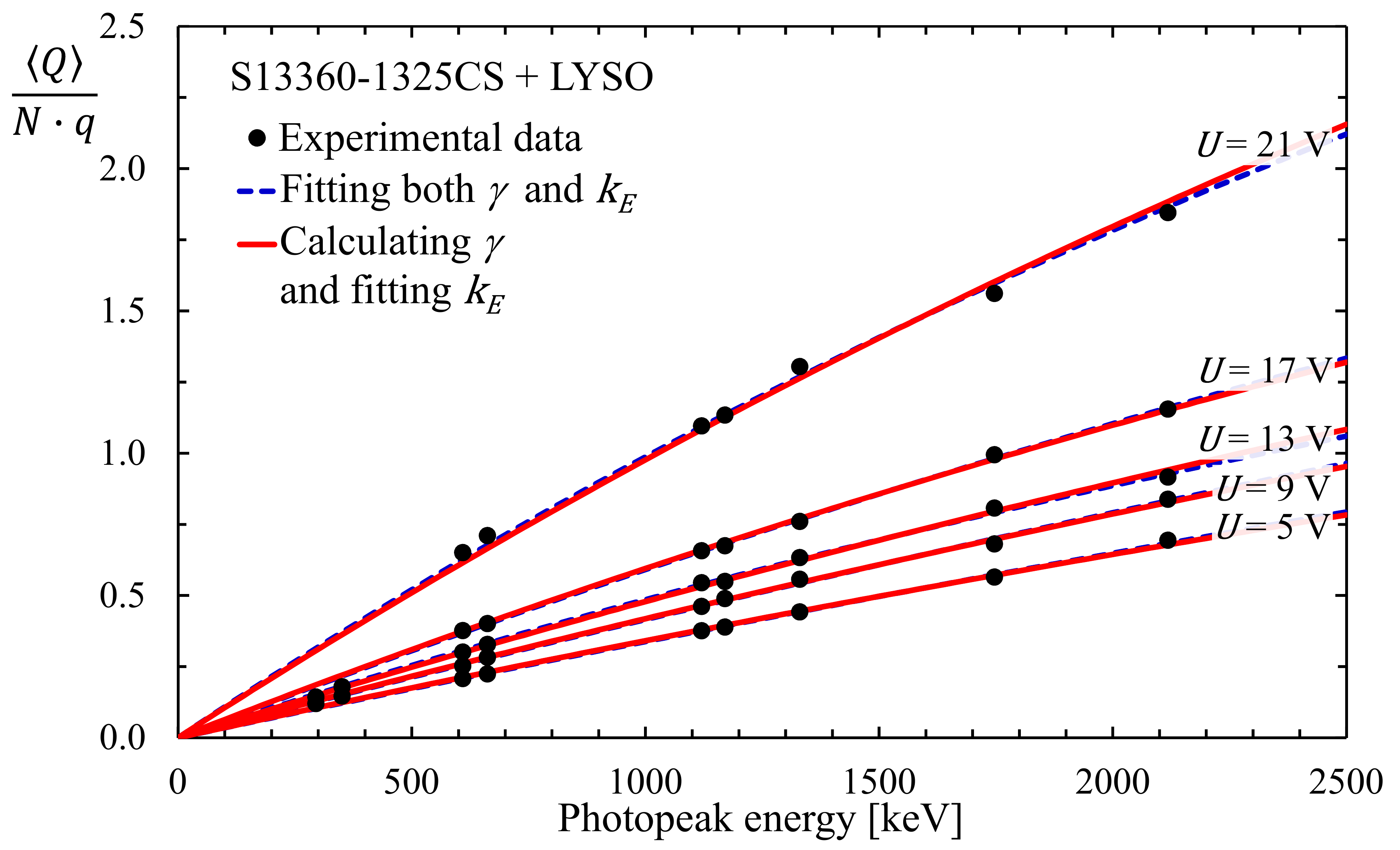}
\includegraphics[width=\linewidth]{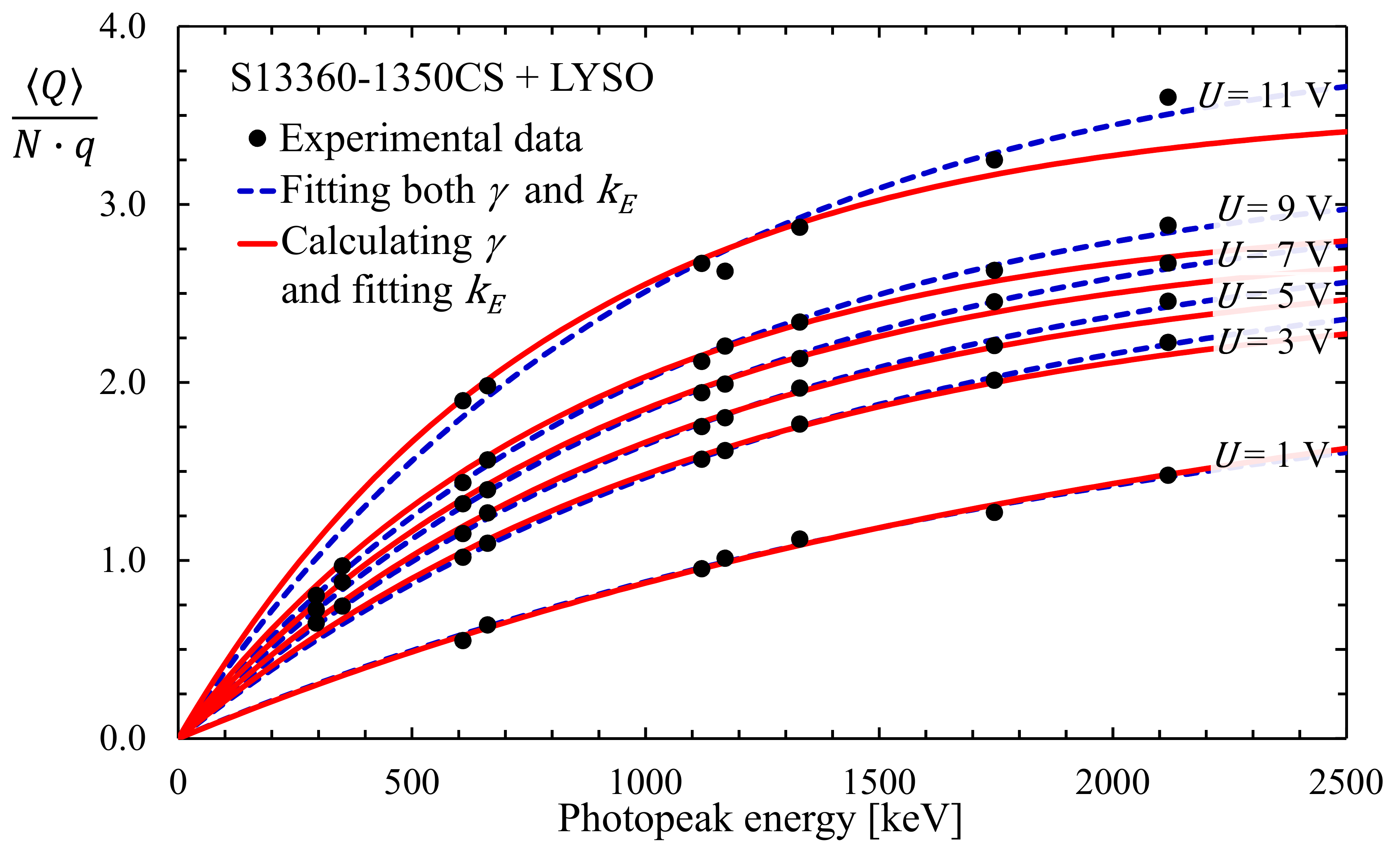}
\caption{Normalized output charge of the two tested SiPMs at several overvoltages for LYSO scintillation pulses as a function of the photopeak energy. Dotted and solid lines represent the two different fits of (\ref{eq:MeanQ_Poiss}) described in the text.}
\label{fig:exponential}
\end{figure}

In Fig. \ref{fig:exponential}, the resolved photopeak positions are represented against the $\gamma$-ray energy $E$. The normalized mean output charge $\langle Q\rangle/(N\cdot q)$ is shown to ease the comparison between both SiPMs. The statistical uncertainties of the photopeak positions are smaller than 5\%, whereas the systematic uncertainty of $\langle Q\rangle/(N\cdot q)$ was estimated to be 10\% due to the uncertainty of $q$. Nonlinearity is more apparent in the 1350 SiPM, because it has less pixels. Similar measurements of the nonlinearity of Hamamatsu SiPMs coupled to LYSO crystals were previously reported in \cite{Niu2012,Pulko2012}.

The number of impinging photons per scintillation pulse is unknown, but it can be assumed to be proportional to $E$. Therefore, the substitution
\begin{equation}
\label{eq:kappa}
(1+c)\cdot n=(1+c)\cdot \eta\cdot Y\cdot E=k_E\cdot E\
\end{equation}
was made in (\ref{eq:MeanQ_Poiss}), where $Y$ is the light yield of the scintillation and $\eta$ is the light collection efficiency, which should be the same for both SiPMs because they have equal area. The coefficient $k_E$ is the only fit parameter and it is basically determined by the data in the linear region in which $\langle Q\rangle\approx\varepsilon\cdot q\cdot k_E\cdot E$.

\begin{figure}[!t]
\centering
\includegraphics[width=\linewidth]{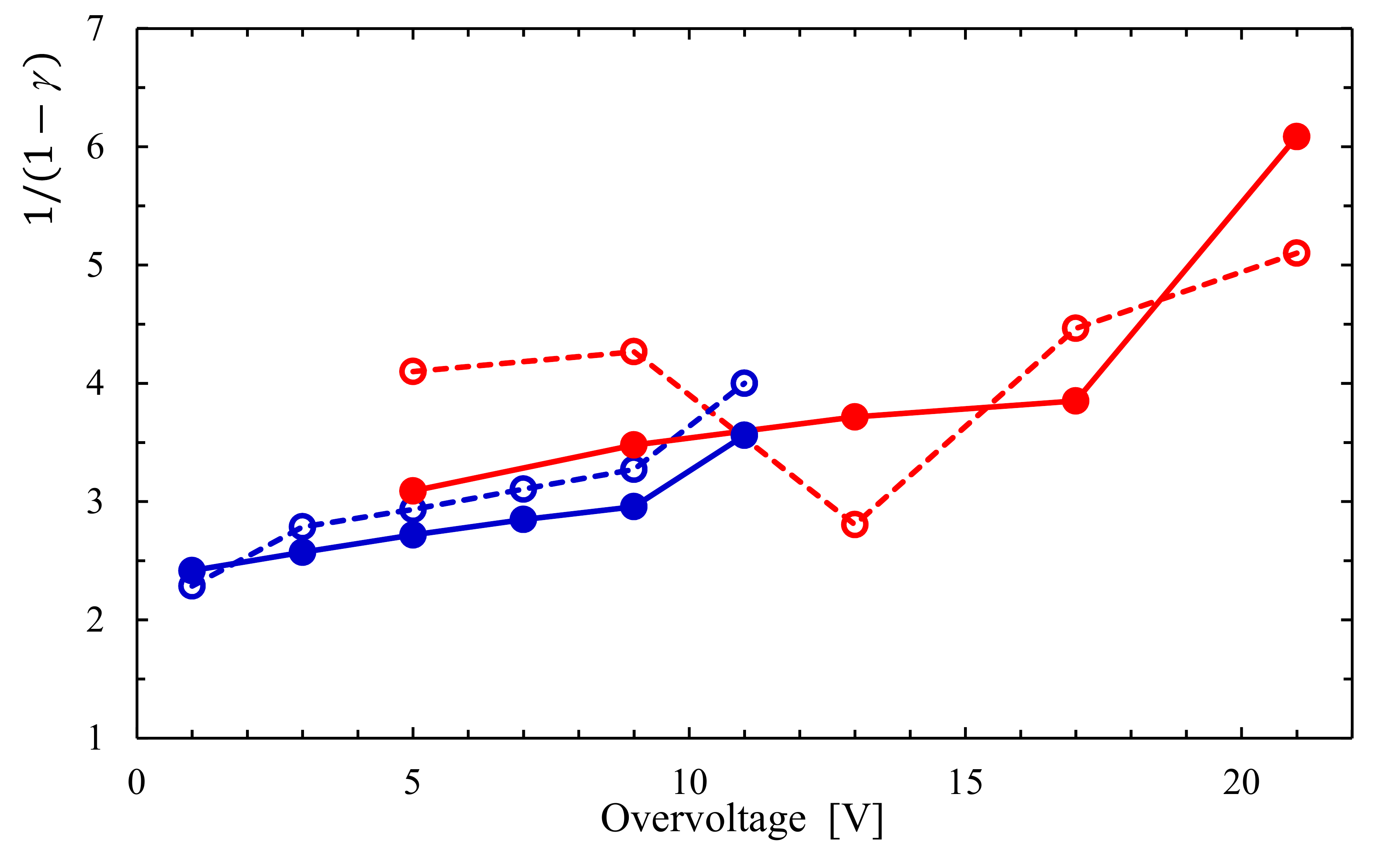}
\includegraphics[width=\linewidth]{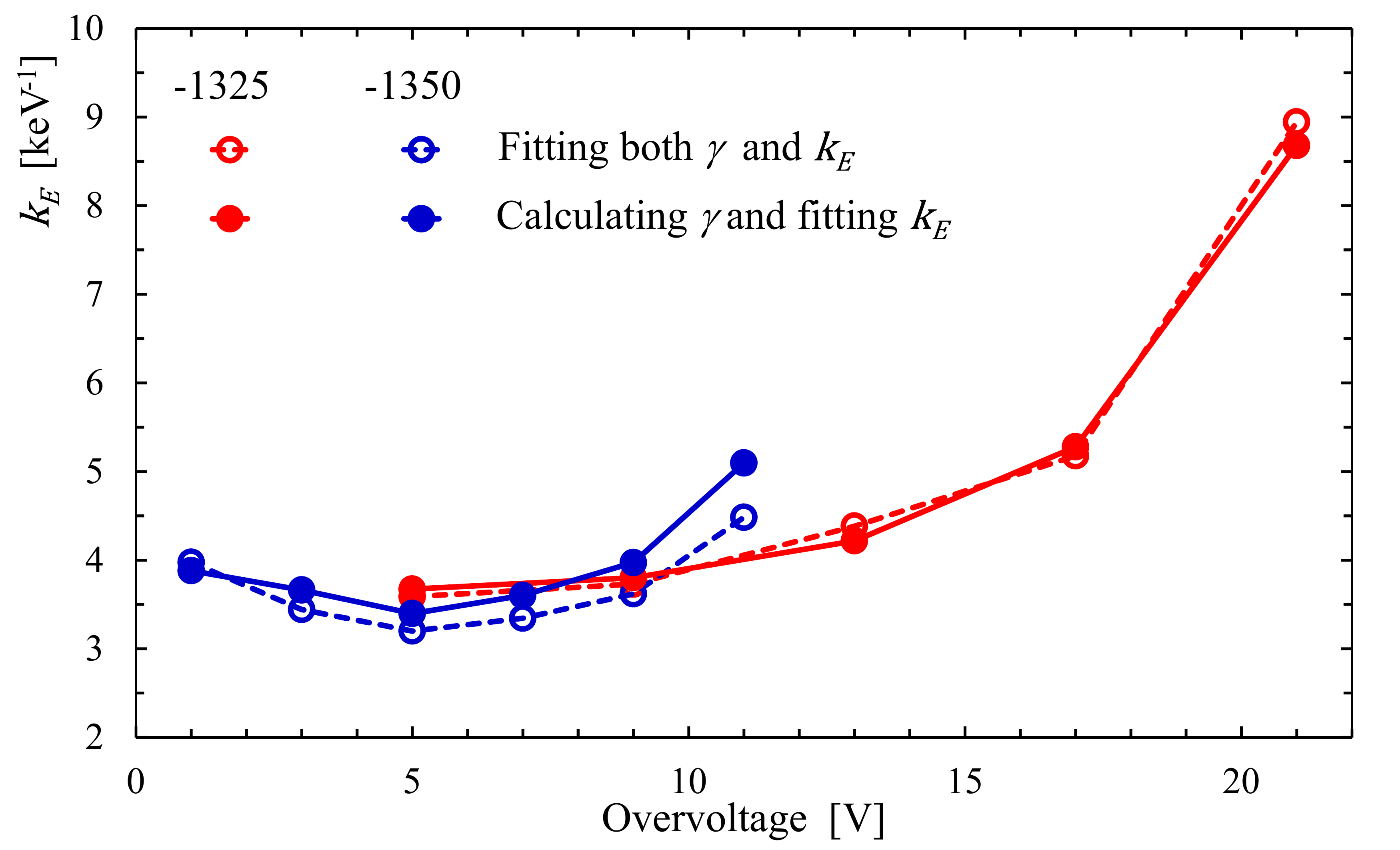}
\caption{Values of $1/(1-\gamma)$ (top) and $k_E$ (bottom) used in the two fits of the model shown in Fig. \ref{fig:exponential}. Filled circles represent the results when $\gamma$ is calculated using the normalized output pulse shape $p(t)$ and $k_E$ is the only free parameter of the fit. Open circles represent the values when both parameters are fitted.}
\label{fig:gamma&kappa}
\end{figure}

For the calculation of the parameter $\gamma$, $p(t)$ was obtained by averaging many output pulses at each $U$ value and normalizing the integral to unity. The recorded pulses were checked to have essentially the same shape regardless of their amplitude. As long as the applied overvoltage was not much higher than the recommended value $U_{\rm op}$, $p(t)$ was well fitted by a double exponential function (\ref{eq:exponential}) with $\tau_1=12$ ns and $\tau_2=45$ ns for the 1325 SiPM, and with $\tau_1=15$ ns and $\tau_2=60$ ns for the 1350 SiPM. However, the pulse length was found to increase rapidly with overvoltage when it exceeded a certain value, namely 20 V for the 1325 SiPM and 10 V for the 1350 SiPM, as illustrated for the latter SiPM in Fig. \ref{fig:pout}. When the overvoltage was increased even further, the signal became very noisy and it was not possible to measure the output pulse charge. This pulse widening coincides with a sudden increase in $\langle Q\rangle/(N\cdot q)$ at high overvoltage (see Fig. \ref{fig:exponential}), which is attributed to the fact that the correlated noise is high enough to produce a self-sustained chain of secondary avalanches. This effect was accounted for by calculating $\gamma$ from (\ref{eq:gamma_exp}) using the fitted values of $\tau_1$ and $\tau_2$ at each overvoltage. Results are represented by filled circles connected by solid lines (in blue for the 1350 SiPM and in red for the 1325 SiPM) in the upper plot of Fig. \ref{fig:gamma&kappa}. Instead of $\gamma$, the saturation level $1/(1-\gamma)$ is shown. Notice that $1/(1-\gamma)$ increases smoothly with overvoltage until the correlated noise starts to become significant.

Solid lines in Fig. \ref{fig:exponential} represent the results from (\ref{eq:MeanQ_Poiss}) when using these calculated values of $\gamma$ and fitting $k_E$ to data. The model describes correctly the SiPM response and its dependence on overvoltage as long as nonlinearity is not very strong. For the 1350 SiPM, the predicted saturation level is about 4\% lower than the measured one, but discrepancies are within the experimental systematic uncertainties.

The fitted values of $k_E$ are represented by filled circles connected by solid lines in the lower plot of Fig. \ref{fig:gamma&kappa}. Statistical uncertainties are smaller than the data point size. At low overvoltage where the correlated noise is very small, $k_E$ is approximately 3.5 keV$^{-1}$ for both SiPMs. Taking $Y=29$ photons/keV and $c=0$ in (\ref{eq:kappa}) results in a light collection efficiency of $\eta=12\%$, which is reasonable considering that the SiPM area covers only 18\% of the crystal surface.

For the 1325 SiPM, $k_E\approx 9$ keV$^{-1}$ at $U=21$ V, which translates into $c\approx 1.5$, that is, the charge contribution of correlated noise is as high as 60\% of the output charge, suggesting that chains of secondary avalanches are produced efficiently at such a high overvoltage. The overall probability of correlated noise is 0.5 at $U=21$ V for this SiPM according to data from \cite{Rosado2015}. Therefore, if no more than one secondary avalanche could be induced by each primary avalanche, then $c$ should be at most 0.5 at this overvoltage.

For the 1350 SiPM, the fitted values of $k_E$ do not show the expected monotonous increase with overvoltage. This may be explained by the fact that $\varepsilon$ was obtained by fitting the approximated expression (\ref{eq:varepsilon}) to data for 450 nm photons, not for the specific emission spectrum of the LYSO crystal.

Tests were also made where $\gamma$ was calculated using the deconvolution of the output pulse signal from the single-photon response function, which can be approximated by an exponential function with decay time of 9 ns for both SiPMs. The main effect of the deconvolution is to make the rising edge of the output pulse faster (i.e., $\tau_1$ decreases down to a few nanoseconds). As a result, the saturation level is reduced by around 9\% and 6\% for the 1325 SiPM and 1350 SiPM, respectively, whereas $k_E$ is basically unchanged.

For consistency, the model was also fitted to the data taking both $\gamma$ and $k_E$ as a free parameters. The best fit is represented by dotted lines in Fig. \ref{fig:exponential}, and the fitted values of $1/(1-\gamma)$ and $k_E$ are shown as open circles connected by dotted lines in Fig. \ref{fig:gamma&kappa}. For the 1325 SiPM, this fit gives almost identical results to the previous one, except for the fact that the uncertainties of the values of $1/(1-\gamma)$ are 20\% or larger, since the nonlinearity is small in this energy range. For the 1350 SiPM, the fitted values of $1/(1-\gamma)$ are slightly higher than the calculated ones, resulting in a better agreement of the model with experimental data.

\begin{figure}[!t]
\centering
\includegraphics[width=\linewidth]{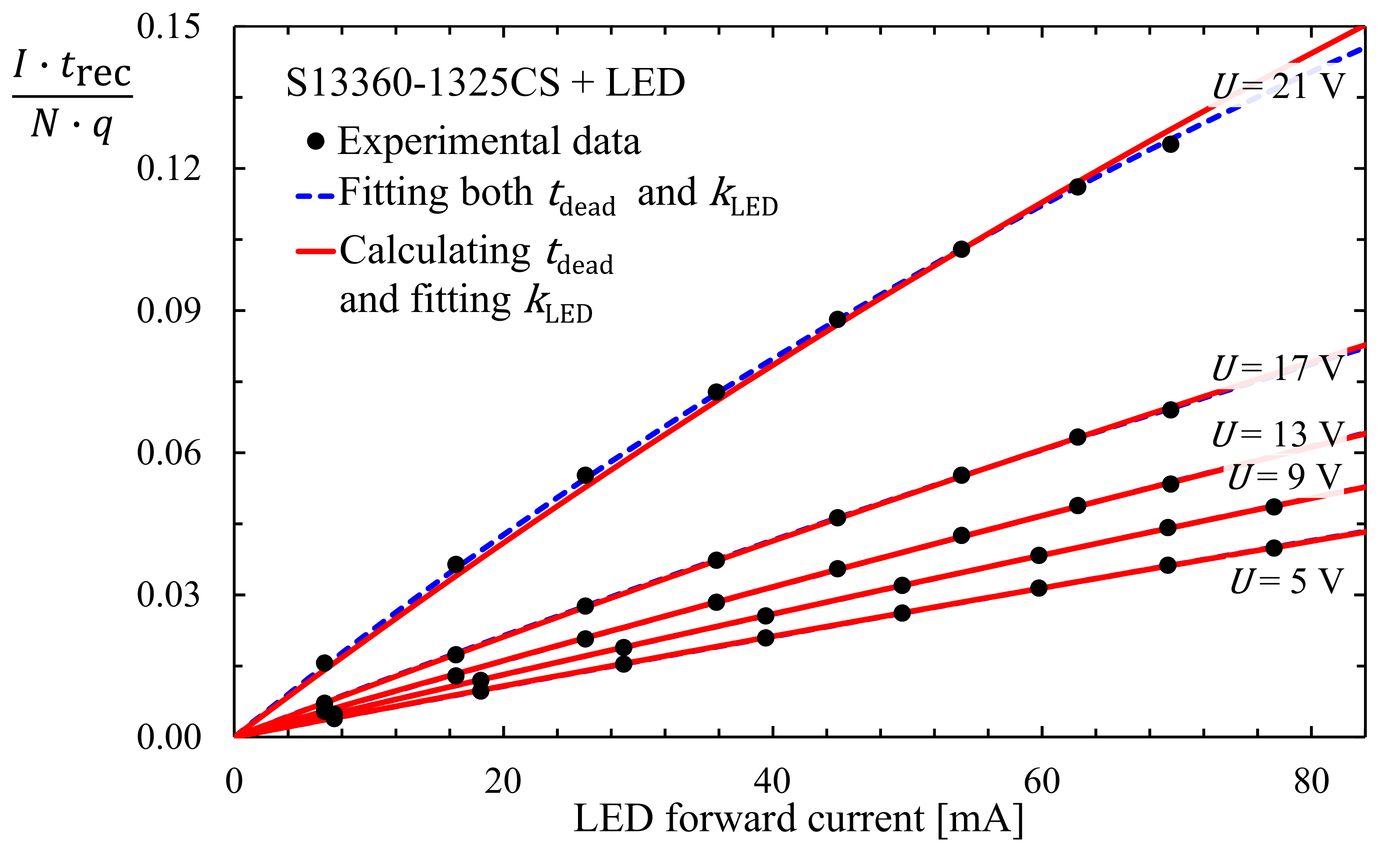}
\includegraphics[width=\linewidth]{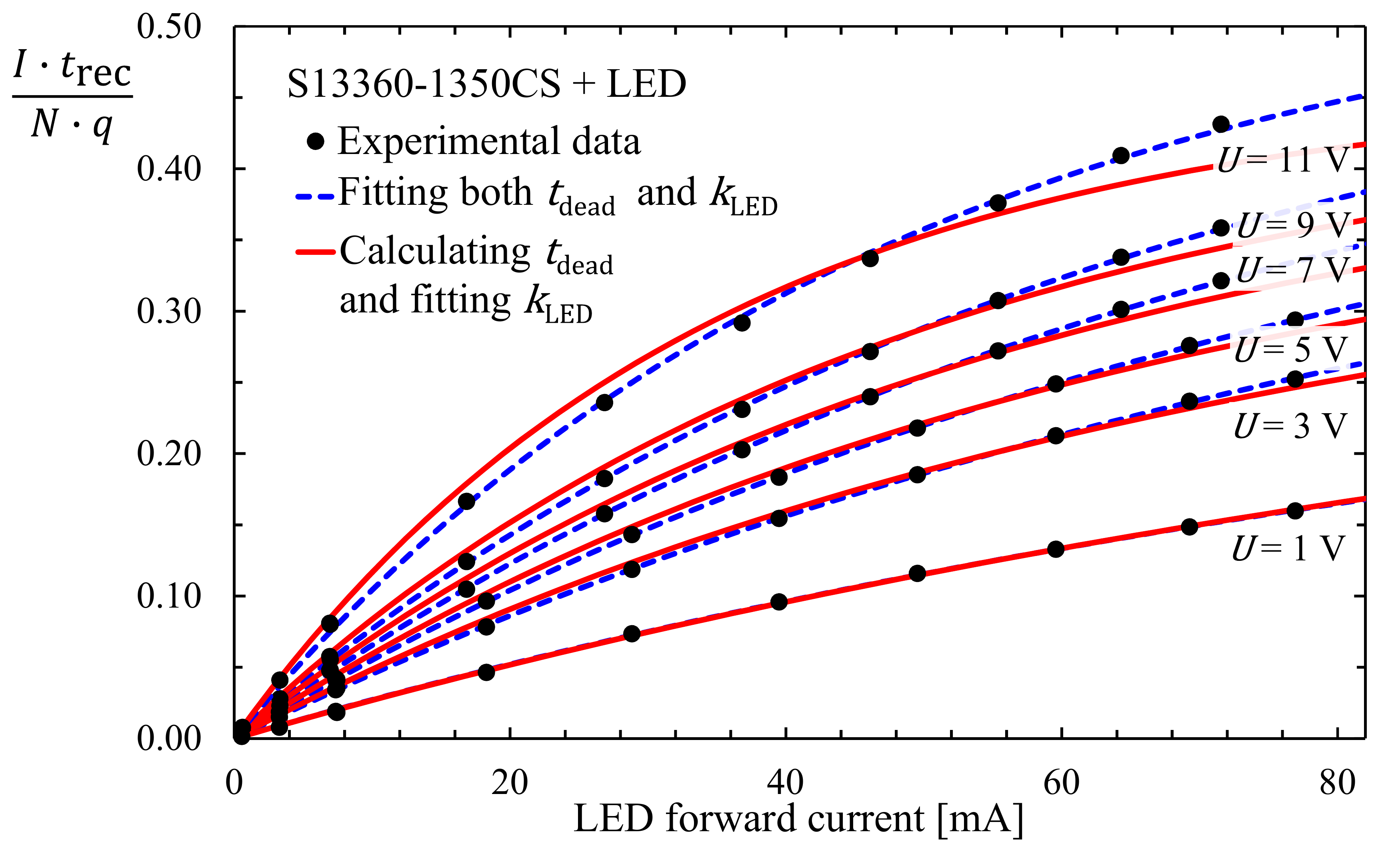}
\caption{Normalized output current of the two tested SiPMs at several overvoltages for continuous light from a LED as a function of the LED forward current. Dotted and solid lines represent the two different fits of (\ref{eq:I_cont}) described in the text.}
\label{fig:LED}
\end{figure}

\subsection{Continuous light}
\label{ssec:LED}

To characterize the SiPM response for continuous light, a green LED (Kingbright L-53GD) was placed at a few centimeters from the SiPM. The mean output current was measured as a function of the LED forward current $I_{\rm LED}$ at different overvoltage values for both SiPMs. Measurements were also performed placing the LED farther away from the SiPM to ensure that it works in the linear region, checking that the light output of the LED was proportional to $I_{\rm LED}$. The uncorrelated noise was found to be negligible in all the measurements.

Results are shown in Fig. \ref{fig:LED}, where the normalized output current $(I\cdot t_{\rm rec})/(N\cdot q)$ is shown to ease the comparison between both SiPMs. The correlated systematic uncertainty was estimated to be 12\% due to uncertainties of $q$ and $t_{\rm rec}$. Notice that the response curves are of a very similar shape to those shown in Fig. \ref{fig:exponential} for scintillation light pulses.

The substitution $(1+c)\cdot r=k_{\rm LED}\cdot I_{\rm LED}$ was made in (\ref{eq:I_cont}), where $k_{\rm LED}$ is a fit parameter that has a similar meaning as the parameter $k_E$ used for scintillation pulses. The saturation level of $(I\cdot t_{\rm rec})/(N\cdot q)$ is $t_{\rm rec}/(2\cdot t_{\rm dead})$, where the parameter $t_{\rm dead}$ was calculated from (\ref{eq:t_dead}). The best fit of (\ref{eq:I_cont}) to the data is represented by solid lines in Fig. \ref{fig:LED}. The model describes approximately the experimental data, although it somewhat underestimates the saturation level for the 1350 SiPM.

\begin{figure}[!t]
\centering
\includegraphics[width=\linewidth]{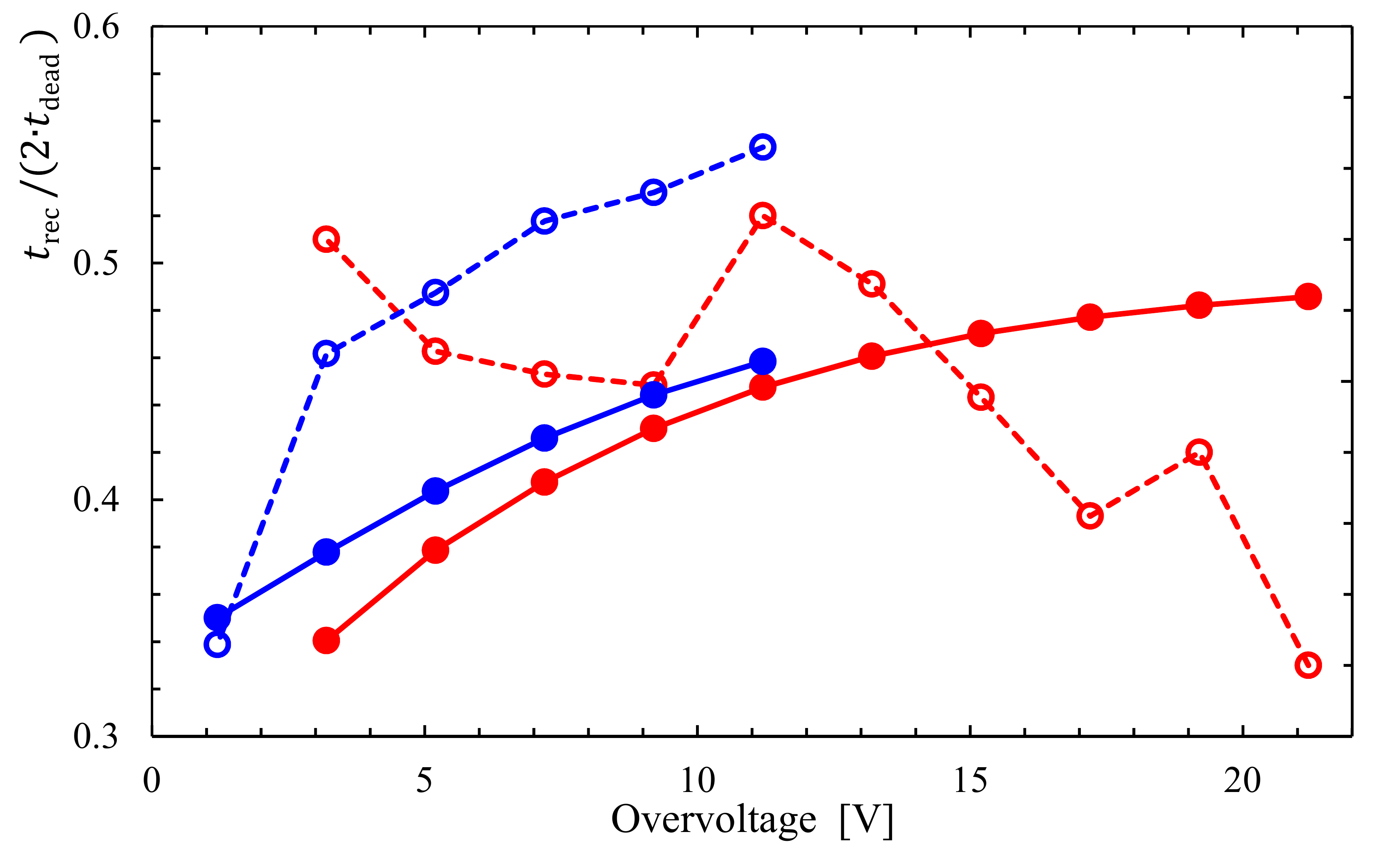}
\includegraphics[width=\linewidth]{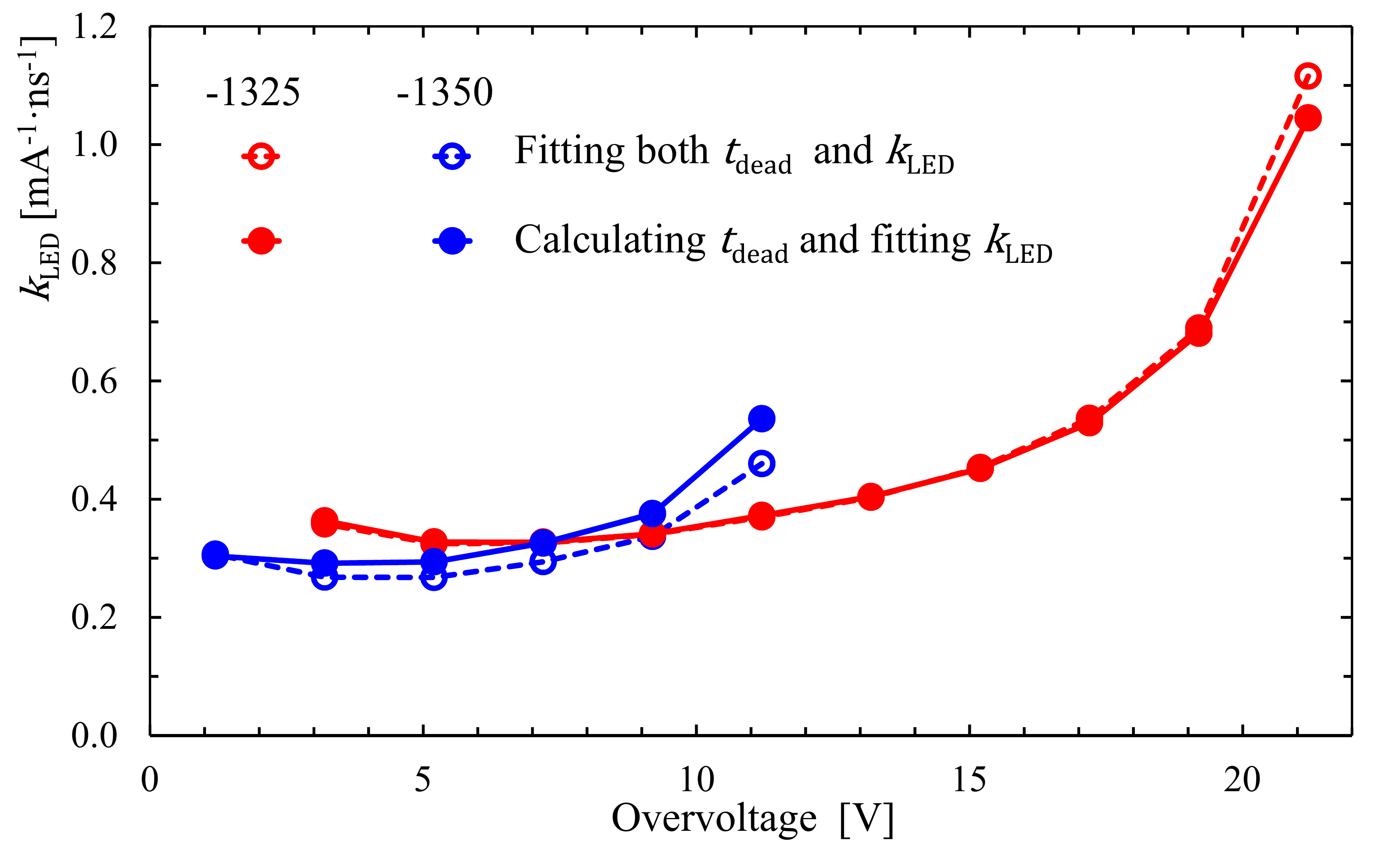}
\caption{Values of $t_{\rm rec}/(2\cdot t_{\rm dead})$ and $k_{\rm LED}$ used in the two fits of the model shown in Fig. \ref{fig:LED}.}
\label{fig:tdead&lambda}
\end{figure}

The calculated values of $t_{\rm rec}/(2\cdot t_{\rm dead})$ and the fitted values of $k_{\rm LED}$ for both SiPMs are represented by filled circles connected by solid lines in Fig. \ref{fig:tdead&lambda}. As mentioned above, $t_{\rm dead}$ approaches $t_{\rm rec}$ when $U$ is large, so that the upper limit of $t_{\rm rec}/(2\cdot t_{\rm dead})$ is $1/2$. Notice that $k_{\rm LED}$ depends on overvoltage in a very similar way to $k_E$ for both SiPMs, which supports that chains of secondary avalanches have an important role in the SiPM response.

The model was also adjusted to data by fitting both $t_{\rm dead}$ and $k_{\rm LED}$. The resulting fit is shown by dotted lines in Fig. \ref{fig:LED} and the fitted values of $t_{\rm rec}/(2\cdot t_{\rm dead})$ and $k_{\rm LED}$ are represented by open circles connected by dotted lines in Fig. \ref{fig:tdead&lambda}. As in the case of the scintillation data, $t_{\rm dead}$ is only poorly determined for the 1325 SiPM because its response is approximately linear in this range. For the 1350 SiPM, the fitted values of $t_{\rm rec}/(2\cdot t_{\rm dead})$ for $U>1$ V are systematically larger than the calculated ones by about 20\%. Nevertheless, both the fitted and the calculated values show the same overvoltage dependence, except for $U=1$ V where the SiPM response is nearly linear.

\section{Conclusions}
\label{sec:conclusions}

A model of the nonlinear response of SiPMs that includes losses of both the avalanche triggering efficiency and the gain of pixels during recovery periods as well as the effects of correlated and uncorrelated noise has been developed. The simple expression (\ref{eq:MeanQ_Poiss}) has been derived, which describes the mean output charge of a SiPM for light pulses of arbitrary shape as long as nonlinear effects and correlated noise are moderate. In addition, the modified expression (\ref{eq:I_cont}) has been obtained for continuous light. The model reduces to the well-known expressions (\ref{eq:InstPulse}) and (\ref{eq:NonParalyzable}) for instantaneous light pulses and for continuous light, respectively, when both correlated and uncorrelated noise are ignored.

Two principal parameters are introduced in the model. First, the parameter $\gamma$ (or $t_{\rm dead}$ in the case of continuous light) gives the saturation level of the SiPM response and it is calculated by a numerical integration knowing the output pulse shape $p(t)$, the pixel recovery time $t_{\rm rec}$ and the photodetection efficiency $\varepsilon$. Simplified expressions for $\gamma$ have also been derived for both rectangular pulses and double exponential pulses. Second, the parameter $c$ represents the relative charge contribution of the correlated noise to the output charge. This parameter is taken as a free parameter that is determined by fitting the model to data in the linear response region of the SiPM. Taking both parameters as a free parameters is nearly equivalent to some previous fitting models used to describe the nonlinear response of SiPM \cite{Grodzicka2015,Niu2012,Pulko2012}, but the present model provides an adequate interpretation of these parameters, including their dependence on both the overvoltage and the output pulse shape.

The model was validated against experimental data of two different SiPMs at moderate nonlinearity using scintillation light pulses from a LYSO crystal. The output pulses were found to be significantly widened when the correlated noise became important, which also caused an increase in the saturation limit of the SiPM response. This effect was accounted for by using the output pulse shape measured at each supplied voltage for the calculation of the parameter $\gamma$. The model describes the data within the experimental uncertainties.

The model also describes adequately the SiPM response for continuous light, although the saturation level was underestimated by about 20\% for one of the SiPMs in this case. More tests are necessary to understand this discrepancy. However, the model describes well the data when both $c$ and $t_{\rm dead}$ are taken as free parameters in the fit. In addition, the overvoltage dependence of the fit parameters is consistent with that obtained for scintillation light pulses.

An adequate parameterization of the chains of secondary avalanches will be useful to provide a theoretical estimation of both the parameter $c$ and the distortion of the output pulse shape. This could be especially relevant for applications where the output pulse height (or a similar parameter) is used instead of the integrated pulse charge.

\appendices

\section{Rectangular pulses}
\label{sec:rectangular}

A normalized rectangular pulse can be expressed as
\begin{equation}
\label{eq:rectangular}
p(t)=\begin{cases}
    \frac{1}{T} & 0<t<T\\
    0 & \rm otherwise\,.
\end{cases}
\end{equation}
Substituting this into (\ref{eq:gamma}) results in
\begin{equation}
\label{eq:gamma_rect}
\gamma=\frac{2}{T^2\cdot\varepsilon\cdot q}\cdot\int_0^{T-t_0}\int_{t_0}^{T-t}\varepsilon_{\rm rec}(t_s)\cdot q_{\rm rec}(t_s)\cdot{\rm d}t_s\cdot{\rm d}t\,.
\end{equation}

For very short pulses such that $T\ll t_{\rm rec}$ and assuming also $t_0\ll t_{\rm rec}$, the first-order Taylor expansions of $\varepsilon_{\rm rec}(t_s)$ and $q_{\rm rec}(t_s)$ about $t_s=0$ can be used in (\ref{eq:gamma_rect}), resulting in
\begin{equation}
\label{eq:gamma_rect_short}
\gamma\approx\frac{\varepsilon_{\rm max}\cdot U}{6\cdot\varepsilon\cdot U_{\rm ch}}\cdot\frac{(T+t_0)\cdot(T-t_0)^3}{T^2\cdot t_{\rm rec}^2}\,.
\end{equation}

On the other hand, for very long pulses, i.e., $t_{\rm rec}\ll T$, the expansion of (\ref{eq:gamma_rect}) to first order in $t_{\rm rec}/T$ and $t_0/T$ gives
\begin{equation}
\label{eq:gamma_rect_long}
\gamma\approx1-\frac{2\cdot t_{\rm dead}}{T}\,,
\end{equation}
where $t_{\rm dead}$ is defined by
\begin{multline}
\label{eq:t_dead}
t_{\rm dead}=t_0+\int_{t_0}^\infty\left[1-\frac{\varepsilon_{\rm rec}(t_s)\cdot q_{\rm rec}(t_s)}{\varepsilon\cdot q}\right]\cdot{\rm d}t_s\\
=t_0+t_{\rm rec}\cdot\left[\frac{\varepsilon_{\rm max}}{\varepsilon}\cdot e^{-t_0/t_{\rm rec}}
-\frac{U_{\rm ch}}{U}\right.\\
+\left.\left(\frac{\varepsilon_{\rm max}}{\varepsilon}-1\right)\cdot\sum_{n=1}^\infty\frac{\left(U/U_{\rm ch}\right)^n}{n\cdot n!}\cdot\left(e^{-t_0/t_{\rm rec}}\right)^n\right]\,.
\end{multline}

\section{Double exponential pulses}
\label{sec:exponential}

In many cases, the output pulse can be fitted by a double exponential function
\begin{equation}
\label{eq:exponential}
p(t)=\frac{1}{\tau_2-\tau_1}\cdot\left(e^{-t/\tau_2}-e^{-t/\tau_1}\right)
\end{equation}
with $t\geq 0$ and $\tau_1<\tau_2$. The rise time and the fall time of the pulse are approximately given by $\tau_1$ and $\tau_2$, respectively, when $\tau_1\ll\tau_2$.

Substituting (\ref{eq:exponential}) into (\ref{eq:gamma}) results in
\begin{equation}
\label{eq:gamma_exp}
\gamma=\frac{\tau_2^2\cdot\gamma_2-\tau_1^2\cdot\gamma_1}{\tau_2^2-\tau_1^2}
\end{equation}
where
\begin{equation}
\label{eq:gamma_i}
\gamma_i=\frac{1}{\tau_i\cdot\varepsilon\cdot q}\cdot\int_{t_0}^\infty e^{-t_s/\tau_i}\cdot\varepsilon_{\rm rec}(t_s)\cdot q_{\rm rec}(t_s)\cdot{\rm d}t_s\;\;\;\;(i=1,\,2)\,.
\end{equation}
Here it has been assumed that the integration time window $T$ is large enough so that $p(t)\approx 0$ for $t>T$.

For $\tau_i\ll t_{\rm rec}$, (\ref{eq:gamma_i}) reduces to
\begin{equation}
\label{eq:gamma_i_short}
\gamma_i\approx \frac{\varepsilon_{\rm max}\cdot U}{\varepsilon\cdot U_{\rm ch}}
\cdot\frac{\tau_i\cdot(2\cdot\tau_i+t_0)}{t_{\rm rec}^2}\cdot e^{-t_0/\tau_i}\,.
\end{equation}

On the other hand, for $t_{\rm rec}\ll\tau_i$, (\ref{eq:gamma_i}) reduces to
\begin{equation}
\label{eq:gamma_i_long}
\gamma_i\approx1-\frac{t_{\rm dead}}{\tau_i}\,,
\end{equation}
where it has been used that $1-[\varepsilon_{\rm rec}(t_s)\cdot q_{\rm rec}(t_s)]/(\varepsilon\cdot q)$ goes to zero much faster than $\exp\left(-t_s/\tau_i\right)$. Therefore, for a very long pulse such that $t_{\rm rec}\ll\tau_1<\tau_2$, one gets
\begin{equation}
\label{eq:gamma_exp_long}
\gamma\approx1-\frac{t_{\rm dead}}{\tau_1+\tau_2}\,.
\end{equation}

The particular case of $\tau_1=0$ in (\ref{eq:exponential}) corresponds to a single exponential function with decay time $\tau=\tau_2$, for which $\gamma$ reduces to $\gamma_2$, which is given by (\ref{eq:gamma_i}).

\section{Corrections for uncorrelated noise}
\label{sec:uncorrelated}

The uncorrelated noise can be readily accounted for by supposing a low rate $r_{\rm bg}$ of \emph{background photons} in addition to those from the light source. This rate $r_{\rm bg}$ can be characterized by measuring the background output charge $\langle Q\rangle_{\rm bg}$ in a given integration time window $T$. Assuming linearity, it is obtained that
\begin{equation}
\label{eq:r_bg}
\langle Q\rangle_{\rm bg}=\varepsilon\cdot r_{\rm bg}\cdot T\cdot q\cdot(1+c)\,,
\end{equation}
where $\varepsilon\cdot r_{\rm bg}$ can be regarded as a continuous rate of background seeds, each one contributing with a mean net charge $q\cdot(1+c)$. It should be clarified that the dark count rate $r_{\rm dc}$, usually given in datasheet specifications, is expected to be slightly larger than $\varepsilon\cdot r_{\rm bg}$, because $r_{\rm dc}$ includes afterpulses surpassing the detection threshold.

The effect of the uncorrelated noise is twofold. In the first place, the effective number of photons $n_{\rm eff}=n+r_{\rm bg}\cdot T$ should be substituted for $n$ in (\ref{eq:MeanQ_Poiss}). In the second place, the effective time distribution $p_{\rm eff}(t)$ including uncorrelated noise is
\begin{equation}
\label{eq:p_eff}
p_{\rm eff}(t)=(1-\chi)\cdot p(t)+\chi\cdot\frac{1}{T}\,,
\end{equation}
where $\chi=(r_{\rm bg}\cdot T)/n_{\rm eff}$ so that $\int_0^T p_{\rm eff}(t)\cdot{\rm d}t=1$. Assuming $\chi\ll 1$, the effective parameter $\gamma_{\rm eff}$ is calculated from
\begin{multline}
\label{eq:gamma_eff}
\gamma_{\rm eff}=(1-2\cdot\chi)\cdot\gamma+\frac{2\cdot\chi}{T\cdot\varepsilon\cdot q}
\cdot\int_0^{T-t_0} \int_{t_0}^{T-t} \left[p(t)\right.\\
\left.+p(t+t_s)\right]
\cdot\varepsilon_{\rm rec}(t_s)\cdot q_{\rm rec}(t_s)\cdot{\rm d}t_s\cdot{\rm d}t\,.
\end{multline}
The first term of the right-hand part of this equation includes only the contribution of seeds produced by the light pulse, where $\gamma$ is calculated from (\ref{eq:gamma}) assuming that $T$ is large enough to contain the full pulse, and the second term arises from interactions of seeds produced by the light pulse with background seeds. To first order, interactions between background seeds can be neglected.

Corrections for uncorrelated noise are only expected to be significant for long light pulses requiring $T\gtrsim 1$ $\mu$s, because $r_{\rm dc}$ is typically lower than 1000 kcps \cite{Hamamatsu}.

\section*{Acknowledgment}
This work was supported by Spanish MINECO under the contract FPA2017-82729-C6-3-R.

\bibliography{IEEEabrv,Bibliography}

\end{document}